# Coordinated Frequency-Constrained Stochastic Economic Dispatch for Integrated Transmission and Distribution System via Distributed Optimization

Ye Tian, *Student Member, IEEE*, Zhengshuo Li, *Senior Member, IEEE*

*Abstract*—When large-scale uncertain centralized and distributed renewable energy sources are connected to a power system, separate dispatching of the transmission power system (TPS) and the active distribution network (ADN) will lower the network security and frequency security of the system. To address these problems, this paper proposes a coordinated frequency-constrained stochastic economic dispatch (CFC-SED) model for an integrated transmission and distribution (ITD) system. In this model, the dynamic frequency security constraints and network security constraints of the ITD system are constructed, and the joint chance constraints are adopted to handle the uncertainty. Then, the control parameters of inverter-based resources, the base point power, and the regulation reserve of all dispatchable resources in the ITD system are jointly optimized for the minimum operating cost. TPS and ADNs can deliver base point power bidirectionally and provide frequency regulation support bidirectionally, which extend the existing reserve assumption in ITD dispatch and enhance the operational security of the ITD system. Moreover, based on the alternating direction of multiplier algorithm, a two-layer distributed optimization framework is proposed to solve the CFC-SED model. Case studies show that the CFC-SED model can fully utilize the potential of multiple regulation resources to improve the security performance of the ITD system, and TPS and ADNs can be coordinated efficiently through the proposed distributed optimization framework.

## Nomenclature

*Indices and sets*

| | |
|---|---|
| $T/D$ | Subscript or sub-subscript markers for distinguishing parameters/variables of the transmission power system (TPS) and active distribution network (ADN) |
| $Tb, Te, Tg, Ti, Tl, Tw$ | Indices of boundary bus, energy storage (ES), thermal unit, bus, transmission line, and dispatchable windfarm (DWF) in TPS |
| $Db, De, Dg, Di, Dl, Dj$ | Indices of boundary bus, ES, distributed generator, bus, transmission line, and dispatchable photovoltaic (DPV) in ADN |
| $E_T, G_T, Line_T, W_T, N_T$ | Sets of ESs, thermal units, transmission lines, DWFs, and bus nodes in TPS |
| $E_D, G_D, Line_D, PV_D, N_D$ | Sets of ESs, distributed generators, transmission lines, DPVs, and bus nodes in ADN |
| $N_B, N_{IBR}$ | Sets of boundary buses connected with ADNs in TPS and inverter-based resources (IBRs) in TPS or ADNs |

*Parameters*

| | |
|---|---|
| $A_P, A_Q, B_P, B_Q$ | Linear coefficient matrices of voltage phase angle and voltage amplitude with respect to nodal injected active/reactive power |
| $F_P, F_Q, F_P^Q, F_Q^Q$ | Linear coefficient matrices of line active/reactive power flow with respect to nodal injected active/reactive power |
| $\beta_m^C, \beta_m^H, \beta_m^D$ | The $m^{th}$ segment piecewise linear fitting coefficients of frequency safety margin |
| $C_{Tg}(\cdot), C_{Dg}(\cdot)$ | Fuel cost functions of generation units $Tg$ and $Dg$ |
| $C_{Tg}^u, C_{Tg}^d, C_{Dg}^u, C_{Dg}^d$ | Cost coefficients of up/down reserves of units $Tg$ and $Dg$ |
| $C_{Te}^{Eu}, C_{Te}^{Ed}, C_{De}^{Eu}, C_{De}^{Ed}$ | Cost coefficients of up/down reserves of ES $Te$ and $De$ |
| $C_{Te}^{loss}, C_{De}^{loss}$ | Cost coefficients of charging/discharging losses of ES $Te$ and $De$ |
| $C_{Tw}^W, C_{Dj}^{PV}$ | Cost coefficients of curtailment of DWF $Tw$ and DPV $Dj$ |
| $d_{Ti}, \tilde{d}_{Ti}, d_{Di}, \tilde{d}_{Di}$ | Forecasted and actual (marked by ~) active demands at buses $Ti$ and $Di$ |
| $\zeta^{max/min}$ | Max/min active power disturbance |
| $q_{Ti}, q_{Di},$ | Forecasted reactive demands at buses $Ti$ and $Di$ |
| $\bar{P}_{Tw}^W, \tilde{P}_{Tw}^W, \bar{P}_{Dj}^{PV}, \tilde{P}_{Dj}^{PV}$ | Forecasted and actual (marked by ~) maximum outputs of DWF $Tw$ and DPV $Dj$ |
| $f_0, \Delta f$ | Normal frequency and frequency variation of the system |
| $\Delta \bar{f}_{rate}, \Delta \bar{f}_{max}, \Delta \bar{f}_{ss}$ | The maximum acceptable absolute value of the rate of change of frequency (RoCoF), maximum frequency deviation, and steady-state frequency deviation |
| $\bar{H}_k, \bar{D}_k$ | Upper bound of virtual inertia and droop coefficient of IBR $k$ |
| $R_k^u, R_k^d$ | Up/down reserves of IBR $k$ |
| $F_{low,Tl}^{max}, S_{flow,Dl}^{max}, S_b$ | Active power capacity of transmission line $l$, and line capacities of transmission line $Dl$ and boundary bus $b$ |
| $S_\theta, S_V$ | Sensitivity matrices of node voltage phase angle and amplitude with respect to node active power variation |
| $V_{un}^{max/min}, \theta_{un}^{max/min}$ | Upper/lower limit of node voltage amplitude and phase angle except slack bus nodes |
| $\Delta V_{un}^{max/min}, \Delta \theta_{un}^{max/min}$ | Max/min variation range of node voltage amplitude and phase angle except slack bus nodes |
| $\Delta t$ | Dispatch time period. |
| $\alpha_g, R_g$ | Adjustable AGC allocation factor and frequency adjustment coefficient of unit $g$ |
| $\delta^V, \delta^L, \delta^{IBR}, \delta_{SFR}^{ITD}$ | Significance level of node voltage violation, power flow overload on transmission line, insufficient up reserves of IBR, and insufficient SFR reserves |

*Variables*

| | |
|---|---|
| $D_k, D_{total}$ | Droop control coefficient of IBR $k$ including DWF, DPV, and ES, and total droop capacity of TPS or ADN |
| $H_k, H_{total}$ | Virtual inertia of IBR $k$ including DWF, DPV, and ES, and total inertia of TPS or ADN |
| $P_{Tb}^B, Q_{Tb}^B, P_{Db}^B, Q_{Db}^B$ | Base point active/reactive power at boundary buses $Tb$ and $Db$ |
| $P_{Te}^E, P_{Te}^{loss}, P_{De}^E, P_{De}^{loss}$ | Charging/discharging power and power loss of ES $Te$ and $De$ |
| $P_{Tg}, Q_{Tg}, P_{Dg}, Q_{Dg}$ | Base point active/reactive power of units $Tg$ and $Dg$ |
| $P_{Tw}^W, Q_{Tw}^W, P_{Dj}^{PV}, Q_{Dj}^{PV}$ | Base point active/reactive power of DWF $Tw$ and DPV $Dj$ |
| $P_{Tl}, Q_{Tl}, \tilde{P}_{Tl}, \tilde{Q}_{Tl}$ $P_{Dl}, Q_{Dl}, \tilde{P}_{Dl}, \tilde{Q}_{Dl}$ | Active/reactive power flow in the base case and uncertainty of line $Tl$ and $Dl$ |

This work was supported by the National Key R&D Program of China under Grant 2022YFB2402900. Y. Tian and Z. Li are with the School of Electrical Engineering, Shandong University, Jinan 250061, China. Zhengshuo Li is the corresponding author (e-mail: zsli@sdu.edu.cn).

| $R_{Tg}^u, R_{Tg}^d, R_{Dg}^u, R_{Dg}^d$ | Up/down reserve capacities of units $Tg$ and $Dg$ |
| --- | --- |
| $R_{Te}^{Eu}, R_{Te}^{Ed}, R_{De}^{Eu}, R_{De}^{Ed}$ | Up/down reserve capacities of units $Te$ and $De$ |
| $R_{D,b}^u, R_{D,b}^d$ | Total up/down regulation reserves of the $b^{th}$ ADN |
| $\Delta \bar{p}_m^{\max}$ | The $m^{th}$ segment frequency security region |
| $\Delta p_b$ | Regulation power from ADN to TPS at boundary bus $b$ |

## I. INTRODUCTION

WHEN large-scale renewable energy is integrated into transmission power systems (TPSs) and active distribution networks (ADNs) [1], the uncertainty throughout the transmission and distribution sectors increases significantly. In such circumstances, if the TPS and ADNs are dispatched separately, various problems will emerge, such as boundary power mismatch between the TPS and ADN and potential line congestion [2],[3]. Therefore, the optimal dispatch of an integrated transmission and distribution (ITD) system has received extensive attention e.g., the researches related to unit commitment and economic dispatch (ED) problems of ITD system [3]-[6], so that the multiple dispatchable resources in TPS and ADN are allocated more reasonably, and, as a result, the reliability of the entire system is improved. In addition, since the TPS and ADN are managed by different operators, various distributed algorithms applied to ITD have also been widely studied, such as alternating direction of multipliers algorithm (ADMM) [5], analytical target cascading [4], and heterogeneous decomposition [3].

Existing studies have mainly focused on the reserve capacity and base point power optimization, without considering the dynamic regulation requirements of the ITD. As the penetration proportion of renewable energy increases, the enhanced power disturbance and the reduced inertia of the system aggravate the risk of frequency irregularities. Thus, it is essential to ensure the frequency security for the coordinated operation of ITD. In the past few years, the importance of considering frequency regulation requirements in ED has been explored [7]–[10]. For example, the authors of [11] studied dynamic ED considering ITD and the automatic regulation effect. As an increasing number of traditional thermal units are replaced by both centralized grid-connected renewable energy in TPSs and distributed grid-connected renewable energy in ADNs, the ED model that only relies on thermal units to regulate frequency cannot reliably deal with real-time power disturbances. This results in **unsatisfactory dynamic frequency performance** (e.g., rate of change of frequency [RoCoF] and maximum frequency deviation) during intra-dispatch periods. Fortunately, changing the control parameters (e.g., virtual inertia and droop coefficient) of inverter-based resources (IBRs) at the ED level can mitigate these issues [12], [13] to improve the dynamic frequency performance during intra-dispatch periods. However, previous studies have only focused on the regulation contribution of centralized grid-connected resources in TPS and have not considered the potential regulation capability of distributed generation sources in ADNs.

To ensure the reliable dynamic frequency performance of an ITD system, it is necessary to develop a frequency-constrained ED for the ITD system. The goals should be to make full use of the regulation resources in ADN as a supplement to TPS (or use the resources in TPS to improve the voltage issues of ADN) and guarantee the optimal economy and safety of the entire system. However, there are several challenges to achieving these goals: *1)* Centralized grid-connected renewable energy in TPS and distributed grid-connected renewable energy in ADN both have obvious uncertainties (e.g., the distributed photovoltaic output steeply declines due to sudden dark clouds in ADN), which will affect the safety of ITD. *2)* Network security problems, such as voltage violation, line congestion, and both active and reactive boundary power mismatch [14], should be avoided when generation resources in ITD (e.g., grid-connected windfarm and distributed photovoltaic) are dispatched to provide frequency regulation support. *3)* TPS and ADNs need to be coordinated in both base point operation and frequency regulation to guarantee the operational safety of the entire ITD system. *4)* The ITD dispatch model, which contains dynamic frequency constraints and joint chance constraints dealing with uncertainty, should be solved in an efficient and distributed pattern. Although studies have been conducted to address some of the above challenges [3], [5], [11], they relied on traditional assumptions, such as regarding ADNs as uncertain loads or adjustable generators on the TPS side. For example, in [11] and [15], ADNs are assumed to be adjustable generators to provide reserve support to TPS, whereas in [5], ADNs are treated as disturbance loads in TPS, and TPS provides reserve support to ADNs, which limits the potential of regulatory resources in the ITD. Moreover, to the best of our knowledge, **there is no relevant research on chance-constrained ITD dispatch considering dynamic frequency constraints, nor on frequency regulation coordination-related models and algorithms**.

To address the aforementioned problems, this paper develops a coordinated frequency-constrained stochastic ED (CFC-SED) model for an ITD system. In the CFC-SED model, joint chance constraints (JCCs) [16] are adopted to handle the uncertainty in TPS and ADNs, and the dynamic frequency security constraints of ITD are constructed reasonably. The new and complex cooperative constraints between TPSs and ADNs are introduced, **which differentiates this work from the existing ITD studies**. Additionally, the base point power and reserve capacity of multiple dispatchable resources as well as the control parameters of IBRs (e.g., dispatchable renewable energy units and energy storage (ES)) are jointly optimized to ensure the optimal economy and safety of the entire ITD system. Finally, a two-layer distributed cooperative optimization framework based on the ADMM algorithm is proposed to solve this CFC-SED model in a distributed pattern. The major contributions of this paper are summarized as follows.

1) A novel CFC-SED model for ITD is proposed, in which the multiple dispatchable resources in ITD are reasonably coordinated considering dynamic frequency constraints, JCCs, network security constraints, and cooperative constraints to improve the dispatch reliability and economy. Compared with existing literature (e.g., [3], [5], [11]), this CFC-SED model allows the TPS and ADNs not only to deliver bidirectional base point power, but also to **provide bidirectional frequency regulation support** when the system power disturbance occurs. Therefore, the frequency regulation ability of ITD is fully exploited, and the operation safety of the entire ITD system is significantly enhanced, which is also verified in the case studies.
2) A two-layer distributed cooperative optimization framework is proposed to solve the CFC-SED model. It allows flexible choices of different algorithms, such as ADMM or other distributed cooperative algorithm, to be adopted in the outer layer to achieve effective distributed cooperation between TPS and ADNs, and the sample average approximation [17] (SAA) or its variant (mix-SAA [18]), to be adopted in the inner layer to handle the chance constraints regarding the TPS and ADNs.

The rest of this paper is organized as follows. Section II articulates the operational structure and formulation of the proposed CFC-SED model. Section III introduces the proposed two-layer distributed solution framework and the algorithm. Section IV presents case studies and analysis. Section V summarizes the conclusions and directions for future work.

## II. Modeling of the CFC-SED model for ITD

### A. Operational Structure of CFC-SED

The operational structure of the CFC-SED model is illustrated in Fig. 1, where the *dispatch results* indicate the optimal base point power, regulation reserve, and inverter control parameter. The dynamic frequency regulation requirements (i.e., regulation reserve, virtual inertia, and droop capacity), network security requirements (e.g., bus node voltage and line power flow security), and the uncertain power in TPS and ADNs are considered in this CFC-SED model. Then the base point power, regulation reserve, and inverter control parameters of the dispatchable resources in TPS, including thermal units, centralized grid-connected windfarm, and ES station, and the dispatchable resources in ADNs, including distributed generation units, distributed photovoltaic, and ES, are jointly optimized for the optimal cooperation of ITD.

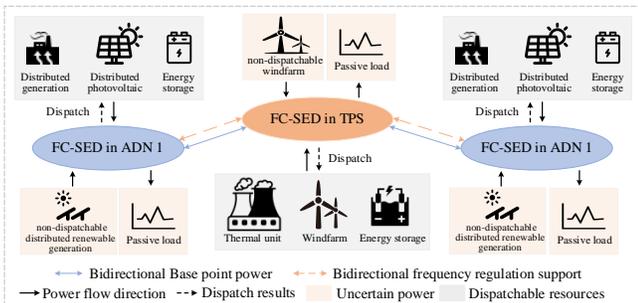

**Fig. 1.** Operational structure of the CFC-SED model.

To achieve the objective of optimal operation and fully utilize the regulation potential of all regulation resources in the ITD, TPS and ADNs are coordinated to optimize base point power and regulation support, as shown in Fig. 1. Bidirectional base point power exchange between TPS and ADNs has been extensively studied in [3], [5], [11], so this paper only focuses on analyzing the bidirectional support between TPS and ADNs during frequency regulation.

Specifically, when a positive disturbance (net load increase or generation reduction) occurs in both TPS and ADNs, there are two possible conditions, as shown in Fig. 2a and 2b: *a)* the ADN can suppress its own disturbance and provides additional regulation reserve or inertia support to TPS; or *b)* the ADN cannot suppress its own disturbance and requires TPS to provide regulation support to ADN. Under these conditions, the boundary power in the positive direction decreases (i.e., the base point power minus the positive power variation) in Fig. 2a because ADN provides regulating power support to TPS; conversely, the boundary power in the positive direction increases in Fig. 2b.

Similarly, when a negative disturbance (net load reduction or generation increase) occurs in both TPS and ADNs, the power variation between TPS and ADNs is shown in Fig. 2c and 2d, respectively. In these figures, the power variation takes a negative value because the power from TPS to ADN will increase when ADN provides regulation support to TPS with a negative disturbance.

When the power disturbances in TPS and ADN are in opposite directions, the power variation between TPS and ADN can also be represented by Fig. 2. For example, if a positive disturbance occurs in TPS and a negative disturbance occurs in ADN, the boundary power from TPS to ADN in the positive direction will decrease, as shown in Fig. 2a.

Finally, notice that the power variation range between TPS and ADNs is subject to the transmission capacity of the boundary bus, as illustrated in Fig. 2e.

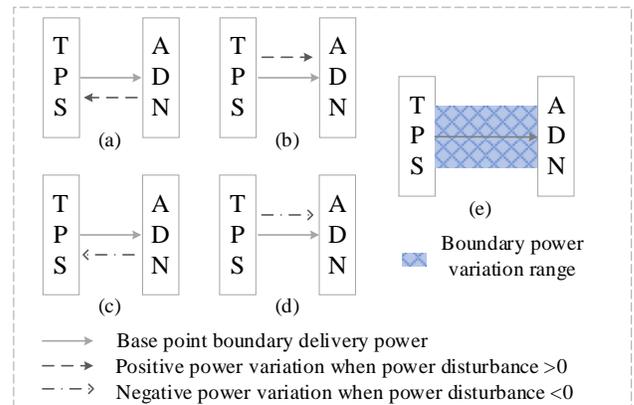

**Fig. 2.** Illustration of the bidirectional regulation support in ITD. In these figures, the solid line indicates the base point boundary power, the arrow points to the positive power direction (if the base point power takes a negative value, it means that the power is reverted from ADN to TPS), and the dotted line indicates the power variation.

Based on the above analysis of new and complex cooperative mechanisms considering frequency regulation, the mathematical formulation of the centralized CFC-SED model is presented below. This CFC-SED model optimizes the control parameters of IBRs, the base point power, and the primary and secondary frequency regulation reserves of all dispatchable resources to achieve the minimum operation cost of ITD.

### B. Constraints in Base Point Operation Case

#### 1) Conventional Operational Constraints

The conventional operational constraints of ITD in the base case have been widely adopted [3], [5], [11] and can be expressed in a compact form, as shown in (1) and (2). The equality constraint in (1) represents active power balance in TPS. The inequality constraint in (1) represents active/reactive base point power limit, reserve capacity, and ramping rate limit constraints of thermal units; active/reactive base point power limit constraints of dispatchable windfarms (DWFs); and base point power, reserve capacity, and state of charge limit constraints for ES in TPS. The equality constraint in (2) represents active/reactive power balance in the $b^{th}$ ADN (the subscript $b$ is omitted for simplicity). The inequality constraint in (2) represents the generation capacity limitation of distributed generators, dispatchable distributed photovoltaics (DPVs), ES, and reactive power compensation in ADN:

$$E_{R_T}x_T+E_{B_T}y_{B_T}=a_T,\ L_{R_T}x_T\geq b_{R_T}, \quad (1)$$

$$E_{R_D}x_D+E_{B_D}y_{B_D}=a_D,\ L_{R_D}x_D\geq b_{R_D}, \quad (2)$$

where vector $x_T$ represents the intra-regional optimized variables in TPS, which include the active/reactive base point power $P_{Tg}, Q_{Tg}$ and up/down regulation reserve $R_{Tg}^u, R_{Tg}^d$ of thermal units; active/reactive base point power $P_{Tw}^W, Q_{Tw}^W$ of DWF; and active base point power $P_{Te}^E$ and up/down regulation reserve $R_{Te}^{Eu}, R_{Te}^{Ed}$ of ES. $y_{B_T}$ denotes the boundary power in TPS, which includes the active/reactive base point power $P_{Tb}^B, Q_{Tb}^B$ at the boundary bus. Similarly, $x_D$ includes $\{P_{Dg}, R_{Dg}^u, R_{Dg}^d, Q_{Dg}\}$, $\{P_{Dj}^{PV}, Q_{Dj}^{PV}\}$, $\{P_{De}^E, R_{De}^{Eu}, R_{De}^{Ed}\}$, and $Q_{rc}^{D,b}$ in ADN. $y_{B_D}$ denotes the boundary power including $\{P_{Db}^B, Q_{Db}^B\}$. $E, L, a, b$ are the coefficient matrices and right-hand vectors of the equality and inequality constraints, which stand for the above conventional operational constraints. The subscripts $R_T, B_T, R_D, B_D$ of the matrices $E, L$ and the vector $b$ indicate that the matrix/vector corresponds to $x_T, y_{B_T}$ and $x_D, y_{B_D}$, respectively. The subscripts $T/D$ are used to distinguish TPS and ADNs.

*2) Network Security Constraints*

To ensure the network security of ITD, the constraints of node voltage amplitude, phase angle, and active power flow for the TPS can be formulated as (3) and (4) based on [19], where $P_T, Q_T$ represent the node injected active/reactive power vector in the base case, and only active power flow is considered because it dominates the apparent power flow of the lines in TPS [20]. $P_T$ consists of $\{P_{Tg}, P_{Tw}^W, P_{Te}^E, P_{Tb}^B, d_{Ti}\}$ and $Q_T$ consists of $\{Q_{Tg}, Q_{Tw}^W, Q_{Tb}^B, q_{Ti}\}$.

$$\theta_{un,T}^{\min}\leq A_{P_T}P_T+A_{Q_T}Q_T\leq\theta_{un,T}^{\max},\ V_{un,T}^{\max}\leq B_{P_T}P_T+B_{Q_T}Q_T\leq V_{un,T}^{\max}. \quad (3)$$

$$P_{Tl}=F_{P_T}P_T+F_{Q_T}Q_T,\ |P_{Tl}|\leq F_{flow,Tl}^{\max},\ \forall Tl\in Line_T. \quad (4)$$

Similarly, the node voltage security constraints and line power flow constraints [21] in ADN can be formulated as (5)–(7), where the node injected active/reactive power $P_D, Q_D$ consist of $\{P_{Dg}, P_{Dj}^{PV}, P_{De}^E, P_{Db}^B, d_{Di}\}$ and $\{Q_{Dg}, Q_{Dj}^{PV}, Q_{Db}^B, q_{Di}\}$, respectively.

$$\theta_{un,D}^{\min}\leq A_{P_D}P_D+A_{Q_D}Q_D\leq\theta_{un,D}^{\max},\ V_{un,D}^{\min}\leq B_{P_D}P_D+B_{Q_D}Q_D\leq V_{un,D}^{\max}. \quad (5)$$

$$P_{Dl}=F_{P_D}P_D+F_{Q_D}Q_D,\ Q_{Dl}=F_{P_D}^Q P_D+F_{Q_D}^Q Q_D. \quad (6)$$

$$(P_{Dl})^2+(Q_{Dl})^2\leq(S_{flow,Dl}^{\max})^2,\ \forall Dl\in Line_D. \quad (7)$$

*3) Boundary Power Constraints*

$$(P_{Tb}^B)^2+(Q_{Tb}^B)^2\leq(S_{Tb})^2,\ (P_{Db}^B)^2+(Q_{Db}^B)^2\leq(S_{Db})^2. \quad (8)$$

$$P_{Tb}^B=P_{Db}^B,\ Q_{Tb}^B=Q_{Db}^B,\ \forall Tb, Db\in N_B. \quad (9)$$

Constraint (8) enforces the base point active/reactive power at the boundary bus to be within the boundary capacity, and $P_{Tb}^B, Q_{Tb}^B$ can take a negative value to indicate that an ADN transfers power to TPS. Constraint (9) is responsible for the matching of boundary base point active/reactive power between TPS and ADNs.

*C. Constraints Under Uncertain Disturbance*

The uncertain disturbances $\zeta_T, \zeta_{D,b}$ of TPS and the $b^{th}$ ADN can be expressed as (10a) and (10b), respectively, where the random variables $\zeta_{Ti}^d, \zeta_{Di}^d$ denote the corresponding uncertain forecast error [16] at bus node $i$. The uncertain disturbance $\zeta^{sys}$ of the entire ITD system can be expressed as (10c), which can also describe a sudden power disturbance (e.g., infeed generation loss, sudden load increase).

$$\tilde{d}_{Ti}=d_{Ti}+\zeta_{Ti}^d,\ \zeta_T=\sum_{Ti\in N_T}\zeta_{Ti}^d. \quad (10a)$$

$$\tilde{d}_{Di}=d_{Di}+\zeta_{Di}^d,\ \zeta_{D,b}=\sum_{Di\in N_{D,b}}\zeta_{Di}^d. \quad (10b)$$

$$\zeta^{sys}=\zeta_T+\sum_{b\in N_B}\zeta_{D,b}. \quad (10c)$$

Different from the traditional ITD dispatch, the CFC-SED model takes the dynamic frequency requirement into account and needs to be able to provide bidirectional frequency regulation support to deal with uncertain disturbance. Thus, it is necessary to structure tailor-made primary frequency regulation (PFR) and secondary frequency regulation (SFR) constraints for ITD dispatch, which are discussed in detail in the following sections.

*1) Dynamic Frequency Response Model*

First, the dynamic frequency response model of the power system with IBRs can be equivalently expressed as (11), whose associated control block diagram is shown in Fig. 3a.

$$\Delta p_G^{eq}-\Delta p_L=2(H_G^{eq}+H_{IBR}^{eq})\frac{d\Delta f}{dt}+(D_O+D_{IBR}^{eq})\Delta f. \quad (11)$$

where $H_G^{eq}$ denotes the aggregated inertia of the thermal units; $D_O$ denotes system damping; $H_{IBR}^{eq}, D_{IBR}^{eq}$, respectively, denote the aggregated virtual inertia and droop coefficient of IBRs; which are aggregated in the same way as [18]; $\Delta p_G^{eq}$ represents the output variation of the aggregated thermal units, which is also a function of $\Delta f$, as shown in Fig. 3a; and $\Delta p_L$ denotes the power disturbance, e.g., $\zeta^{sys}$.

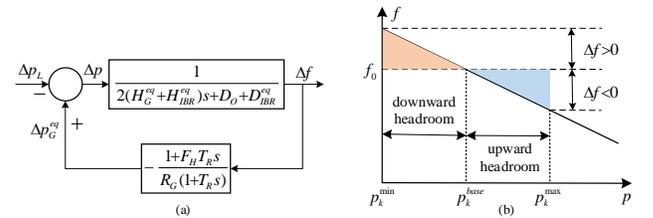

**Fig. 3.** (a) Control block diagram of the equivalent system frequency response model, where $R_G, F_H, T_R$ represent the aggregated droop coefficient, the fraction of power generated by the high-pressure turbine, and the reheat time of thermal units. (b) Upward and downward regulation headroom for IBRs [18].

Then, three representative indices [22] to quantify the dynamic frequency performance of PFR can be expressed as (12a)–(12c), where $F_{un}(H_{IBR}^{eq}, D_{IBR}^{eq})$ represents the nonlinear frequency safety margin of the maximum frequency deviation, and the related derivation is detailed in [23].

**Maximum RoCoF:** $\quad |\frac{-\Delta p_L}{2(H_G^{eq}+H_{IBR}^{eq})}|\leq\Delta\bar{f}_{rate}. \quad (12a)$

**Maximum frequency deviation:** $\left|\frac{-\Delta p_L}{F_{un}(H_{IBR}^{eq},D_{IBR}^{eq})}\right| \leq \Delta \bar{f}_{\max}$. (12b)

**Steady-state frequency deviation:** $\left|\frac{-\Delta p_L}{D_O+D_{IBR}^{eq}+R_G^{-1}}\right| \leq \Delta \bar{f}_{ss}$. (12c)

In addition, the virtual inertia and droop capacity of IBR are realized according to the control behavior shown in (13), where $\Delta p_k$ denotes the output power variation of IBR. Taking only droop control as an example, the relationship between output variation and frequency deviation is shown in Fig. 3b, which requires IBR to have sufficient regulation headroom to reliably provide frequency regulation support. Therefore, the regulation reserve constraints (14) need to be satisfied. To avoid the nonlinearity caused by $d\Delta f/dt$ and $\Delta f$, which are nonlinear functions about control variables $H_k, D_k$, the more conservative reserve constraints (15) are adopted, as in [18].

$$\Delta p_k = -(2H_k d\Delta f/dt + D_k \Delta f), k \in N_{IBR}. \quad (13)$$

**Upward headroom:** $R_k^u \geq |2H_k d\Delta f/dt + D_k \Delta f|, \Delta f < 0, k \in N_{IBR}$. (14a)

**Downward headroom:** $R_k^d \geq |2H_k d\Delta f/dt + D_k \Delta f|, \Delta f > 0, k \in N_{IBR}$. (14b)

$$\begin{aligned} R_k^u \geq 2H_k \Delta \bar{f}_{rate} + D_k \Delta \bar{f}_{\max}, \forall k \in N_{IBR} \\ R_k^d \geq 2H_k \Delta \bar{f}_{rate} + D_k \Delta \bar{f}_{\max}, \forall k \in N_{IBR} \end{aligned} \quad (15)$$

In summary, to ensure satisfactory dynamic frequency performance of the system, the PFR-related constraints (12) and (15) need to be incorporated into the CFC-SED model.

*2) PFR-Related Constraints for ITD*

$$H_T^{total} + \sum_{b \in N_B} H_{D,b}^{total} \geq |\zeta^{sys}|/2\Delta \bar{f}_{rate}. \quad (16a)$$

$$D_T^{total} + \sum_{b \in N_B} D_{D,b}^{total} \geq |\zeta^{sys}|/\Delta \bar{f}_{ss}. \quad (16b)$$

$$\Delta \bar{p}_{T,m}^{\max} + \sum_{b \in N_B} \Delta \bar{p}_{D,b,m}^{\max} \geq |\zeta^{sys}|, \forall m. \quad (16c)$$

$$\Delta \bar{p}_{T,m}^{\max} = \Delta \bar{f}_{\max} (\beta_{T,m}^C + \beta_{T,m}^H H_T^{total} + \beta_{T,m}^D D_T^{total}). \quad (17a)$$

$$\Delta \bar{p}_{D,b,m}^{\max} = \Delta \bar{f}_{\max} (\beta_{D,b,m}^C + \beta_{D,b,m}^H H_{D,b}^{total} + \beta_{D,b,m}^D D_{D,b}^{total}). \quad (17b)$$

$$0 \leq H_k \leq \bar{H}_k, 0 \leq D_k \leq \bar{D}_k, k \in \{W_T, PV_D, E_T, E_D\}. \quad (17c)$$

As mentioned previously, the linear RoCoF constraint, maximum frequency deviation, and quasi-steady-state frequency deviation constraints [12] in ITD can be formulated as (16a), (16b), and (16c), respectively, where (16c) is expressed through a set of linearized segment frequency security margins [23] of TPS and ADNs, which are shown in (17a) and (17b), respectively, and the detailed derivation is shown in [23]. Constraint (17c) represents the variable range of the virtual inertia and droop coefficients of DWF or DPV and ES. Further, since the uncertainty distributions of $\zeta_T, \zeta_{D,b}$ are independent from each other, constraint (16) can be transformed into (18), which is physically interpreted as letting the dynamic frequency performance of PFR be guaranteed for the worst case.

$$H_T^{total} + \sum_{b \in N_B} H_{D,b}^{total} \geq \frac{1}{2\Delta \bar{f}_{rate}} \max\{\zeta_T^{\max} + \sum_{b \in N_B} \zeta_{D,b}^{\max}, -\zeta_T^{\min} - \sum_{b \in N_B} \zeta_{D,b}^{\min}\}. \quad (18a)$$

$$D_T^{total} + \sum_{b \in N_B} D_{D,b}^{total} \geq \frac{1}{\Delta \bar{f}_{ss}} \max\{\zeta_T^{\max} + \sum_{b \in N_B} \zeta_{D,b}^{\max}, -\zeta_T^{\min} - \sum_{b \in N_B} \zeta_{D,b}^{\min}\}. \quad (18b)$$

$$\Delta \bar{p}_{T,m}^{\max} + \sum_{b \in N_B} \Delta \bar{p}_{D,b,m}^{\max} \geq \max\{\zeta_T^{\max} + \sum_{b \in N_B} \zeta_{D,b}^{\max}, -\zeta_T^{\min} - \sum_{b \in N_B} \zeta_{D,b}^{\min}\}, \forall m. \quad (18c)$$

Moreover, to ensure the sufficient regulation ability of ITD, the regulation reserve constraints need to be involved in the CFC-SED model according to (15). Since the maximum output of DWF is uncertain, the upward reserve constraint of DWF is modeled as JCC (19) with the confidence level being $1-\delta_T^{IBR}$ to enhance dispatch reliability, where $\mathbb{P}_{\tilde{P}_{Tw}^W}$ is the probability distribution with respect to the random variable $\tilde{P}_{Tw}^W$. The up/down PFR reserve constraints of ESs and thermal units are formulated as deterministic constraints (20) and (21), respectively, as in [24]. The PFR reserve constraints of all regulation resources in ADN are consistent with TPS and are no longer expressed to avoid repetitiveness.

$$\mathbb{P}_{\tilde{P}_{Tw}^W}\{\tilde{P}_{Tw}^W - P_{Tw}^W \geq 2H_{Tw}\Delta \bar{f}_{rate} + D_{Tw}\Delta \bar{f}_{\max}, \forall Tw\} \geq 1-\delta_T^{IBR}. \quad (19)$$

$$\begin{aligned} R_{Te}^{Eu} \geq 2H_{Te}\Delta \bar{f}_{rate} + D_{Te}\Delta \bar{f}_{\max} \\ R_{Te}^{Ed} \geq 2H_{Te}\Delta \bar{f}_{rate} + D_{Te}\Delta \bar{f}_{\max} \end{aligned}. \quad (20)$$

$$R_{Tg}^u \geq \Delta \bar{f}_{ss}/R_{Tg}, R_{Tg}^d \geq \Delta \bar{f}_{ss}/R_{Tg}. \quad (21)$$

*3) SFR-Related Constraints*

SFR is used to mitigate the power disturbance and restore the frequency to the normal value, so multiple regulation resources also need to have sufficient SFR reserves. The SFR reserve constraints for the entire ITD system can be formulated as chance constraint (22) with the confidence level being $1-\delta_{SFR}^{ITD}$ to make safer decisions under uncertain disturbances, and only thermal units are involved in SFR [18]. However, (22) is difficult to solve because the uncertainty distributions of $\zeta^T, \zeta_b^D$ are independent from each other. For example, when uncertainty information is depicted using the scenario sampling method [17], (22) would contain $n_T(n_D)^{N_B}$ scenarios, causing a dimensional disaster, where $n_T, n_D$ represent the number of uncertain scenarios in TPS and ADNs. Thus, a conservative but safe conversion method is adopted here, as shown in (23)–(24).

$$\mathbb{P}_{\zeta_T,\zeta_{D,b}}\left\{-\sum_{Tg \in G_T} R_{Tg}^d - \sum_{b \in N_B} R_{D,b}^d \leq \zeta_T + \sum_{b \in N_B} \zeta_{D,b} \leq \sum_{Tg \in G_T} R_{Tg}^u + \sum_{b \in N_B} R_{D,b}^u\right\} \geq 1-\delta_{SFR}^{ITD}. \quad (22)$$

By shifting $\zeta_{D,b}$ to both sides of the inequality in (22), the up/down regulation power $\Delta p_b^+, \Delta p_b^-$ available from ADN at boundary bus *b* can be obtained as (23a), while (23b) indicates the upper/lower bound of the corresponding regulation power range. Then, to ensure sufficient SFR reserve for the entire system despite the minimum regulated power provided by ADNs, the SFR reserve constraint can be constructed as (24) to replace (22).

$$\Delta p_b^+ = R_{D,b}^u - \zeta_{D,b}, \quad \Delta p_b^- = R_{D,b}^d + \zeta_{D,b}. \quad (23a)$$

$$\begin{aligned} \Delta p_b^{+,\max} = R_{D,b}^u - \zeta_{D,b}^{\min}, \Delta p_b^{+,\min} = R_{D,b}^u - \zeta_{D,b}^{\max} \\ \Delta p_b^{-,\max} = R_{D,b}^d + \zeta_{D,b}^{\max}, \Delta p_b^{-,\min} = R_{D,b}^d + \zeta_{D,b}^{\min} \end{aligned}. \quad (23b)$$

$$\mathbb{P}_{\zeta_T}\left\{\begin{aligned} \zeta_T \leq \sum_{Tg \in G_T} R_{Tg}^u + \sum_{b \in N_B}(R_{D,b}^u - \zeta_{D,b}^{\max}), \zeta_T \geq 0 \\ -\sum_{Tg \in G_T} R_{Tg}^d - \sum_{b \in N_B}(R_{D,b}^d + \zeta_{D,b}^{\min}) \leq \zeta_T, \zeta_T < 0 \end{aligned}\right\} \geq 1-\delta_{SFR}^{ITD}. \quad (24)$$

Thus far, the frequency-related constraints for the ITD system have been established. During the PFR process, TPS and ADNs jointly provide inertia and droop capacity support to ensure the PFR-related dynamic frequency performance of ITD; during the SFR process, TPS and ADN can provide regulation reserve bidirectionally according to the actual disturbance. For example, when $\Delta p_b^{+,\min} > 0$, ADN satisfies its own disturbance and provides additional reserve support to TPS. In contrast, when $\Delta p_b^{+,\min} < 0$, the TPS is needed to provide regulation reserve to ADN.

*4) Network Security Constraints Under Uncertainty*

When an uncertain disturbance occurs, the successive PFR and SFR processes will cause changes in line power flow and node voltages and may endanger the network security. Hence, the network security should also be imposed to secure the ITD in the PFR and SFR processes.

*a) Nodal Voltage Security Constraints Associated with PFR*

Due to the short PFR process [22], we assume the effect of PFR on the transmission line flow is negligible, while the node voltage security constraints (27) during the PFR process are considered. Specifically, constraints (25) and (26) represent the maximum PFR regulation power of IBRs and thermal units, respectively. Constraint (27) enforces the voltage phase and magnitude changes caused by the maximum PFR regulation power within the limitation, where the maximum nodal injected power variation $\Delta \overline{p}_{T/D}^{PFR}$ is composed of $\Delta \overline{p}_k^{PFR}$ and $\Delta \overline{p}_g^{PFR}$. To avoid duplication, subscripts $T/D$ are used to distinguish constraints in TPS and ADN.

$$\Delta \overline{p}_k^{PFR} = 2H_k \Delta \overline{f}_{rate} + D_k \Delta \overline{f}_{\max}, k \in \{W_T, E_T, PV_D, E_D\}. \quad (25)$$

$$\Delta \overline{p}_g^{PFR} \geq \Delta \overline{f}_{\max} / R_g, g \in [G_T, G_D]. \quad (26)$$

$$\Delta \boldsymbol{\theta}_{un,T/D}^{\min} \leq \boldsymbol{S}_\theta \Delta \boldsymbol{p}_{T/D}^{PFR} \leq \Delta \boldsymbol{\theta}_{un,T/D}^{\max}, \Delta \boldsymbol{V}_{un,T/D}^{\min} \leq \boldsymbol{S}_V \Delta \boldsymbol{p}_{T/D}^{PFR} \leq \Delta \boldsymbol{V}_{un,T/D}^{\max}. \quad (27)$$

*b) Line Power Flow and Node Voltage Security Constraints Associated with SFR*

In the SFR process of TPS, the boundary bus power $\tilde{P}_{Tb}^B$ is the base point boundary power minus the power variation from ADN to TPS at the boundary bus, as shown in (28a), and the output power $\tilde{P}_{Tg}$ of thermal unit $Tg$ can be expressed as (28b) according to (22)–(24). Thus, the uncertain node active power injection $\tilde{\boldsymbol{P}}_T$ is composed of $\{\tilde{P}_{Tg}, P_{Tw}^W, P_{Te}^E, \tilde{P}_{Tb}^B, \tilde{d}_{Ti}\}$, and node reactive power injection $\tilde{\boldsymbol{Q}}_T$ is composed of $\{Q_{Tg}, Q_{Tw}^W, Q_{Tb}^B, q_{Ti}\}$, where the reactive power demand is assumed to be constant [14]. Thus, the voltage security constraints of TPS can be formulated as (29), where the JCC is adopted to improve dispatch reliability under uncertainty. The line active power flow under uncertainty can be expressed as (30a), where $\boldsymbol{F}_{Tl}^+, \boldsymbol{F}_{Tl}^-$ denote the positive and negative coefficient matrices of $\Delta p_b$, respectively, and $\boldsymbol{F}_{Tl}^{rm}$ denotes the residual component. From a safe conservative perspective, the possible max/min line active power flow $\tilde{P}_{Tl}^{\max}, \tilde{P}_{Tl}^{\min}$ can be obtained by relaxing $\Delta p_b$ to the upper/lower boundaries according to the positive/negative coefficients of $\Delta p_b$, respectively. Thus, the line power flow constraints under uncertainty in TPS can be constructed as JCC (30c) with the confidence level $1-\delta_T^L$ to ensure line security.

$$\tilde{P}_{Tb}^B = P_{Tb}^B - \Delta p_b, \Delta p_b \in [-\Delta p_b^{-,\max}, \Delta p_b^{+,\max}]. \quad (28a)$$

$$\tilde{P}_{Tg} = P_{Tg} + \alpha_{Tg}(\zeta_T - \sum_{b \in N_B} \Delta p_b). \quad (28b)$$

$$\mathbb{P}_{\zeta_{Ti}^d} \left\{ \begin{array}{l} \boldsymbol{\theta}_{un,T}^{\min} \leq \boldsymbol{A}_{P_T} \tilde{\boldsymbol{P}}_T + \boldsymbol{A}_{Q_T} \tilde{\boldsymbol{Q}}_T \leq \boldsymbol{\theta}_{un,T}^{\max} \\ \boldsymbol{V}_{un,T}^{\min} \leq \boldsymbol{B}_{P_T} \tilde{\boldsymbol{P}}_T + \boldsymbol{B}_{Q_T} \tilde{\boldsymbol{Q}}_T \leq \boldsymbol{V}_{un,T}^{\max} \end{array} \right\} \geq 1 - \delta_T^V. \quad (29)$$

$$\tilde{\boldsymbol{P}}_{Tl} = \boldsymbol{F}_{P_T} \tilde{\boldsymbol{P}}_T + \boldsymbol{F}_{Q_T} \tilde{\boldsymbol{Q}}_T = \boldsymbol{F}_{Tl}^+ \cdot \Delta p_b + \boldsymbol{F}_{Tl}^- \cdot \Delta p_b + \boldsymbol{F}_{Tl}^{rm}. \quad (30a)$$

$$\tilde{\boldsymbol{P}}_{Tl}^{\max} = \boldsymbol{F}_{Tl}^+ \cdot \Delta p_b^{+,\max} + \boldsymbol{F}_{Tl}^- \cdot (-\Delta p_b^{-,\max}) + \boldsymbol{F}_{Tl}^{rm}$$
$$\tilde{\boldsymbol{P}}_{Tl}^{\min} = \boldsymbol{F}_{Tl}^+ \cdot (-\Delta p_b^{-,\max}) + \boldsymbol{F}_{Tl}^- \cdot \Delta p_b^{+,\max} + \boldsymbol{F}_{Tl}^{rm}. \quad (30b)$$

$$\mathbb{P}_{\zeta_{Ti}^d} \left\{ |\tilde{P}_{Tl}^{\min}| \leq F_{low,Tl}^{\max}, |\tilde{P}_{Tl}^{\max}| \leq F_{low,Tl}^{\max}, \forall Tl \in Line_T \right\} \geq 1 - \delta_T^L. \quad (30c)$$

Similar to TPS, the boundary bus power and output power of a distributed generator of ADNs in the SFR process are shown in (31a), and the uncertain node active and reactive power injections $\tilde{\boldsymbol{P}}_D, \tilde{\boldsymbol{Q}}_D$ consist of $\{\tilde{P}_{Dg}, P_{Dj}^{PV}, P_{De}^E, \tilde{P}_{Db}^B, \tilde{d}_{Di}\}$ and $\{Q_{Dg}, Q_{Dj}^{PV}, Q_{Db}^B, \tilde{q}_{Di}\}$, respectively. The line power flow in ADN is shown in (31b), and the voltage security constraint and conservative line power flow constraint can be expressed as (32) and (33), respectively:

$$\tilde{P}_{Db}^B = P_{Db}^B - \Delta p_b, \tilde{P}_{Dg} = P_{Dg} + \alpha_{Dg}(\zeta_{D,b} + \Delta p_b), \quad (31a)$$

$$\tilde{\boldsymbol{P}}_{Dl} = \boldsymbol{F}_{P_D} \tilde{\boldsymbol{P}}_D + \boldsymbol{F}_{Q_D} \tilde{\boldsymbol{Q}}_D, \tilde{\boldsymbol{Q}}_{Dl} = \boldsymbol{F}_{P_D}^Q \tilde{\boldsymbol{P}}_D + \boldsymbol{F}_{Q_D}^Q \tilde{\boldsymbol{Q}}_D, \quad (31b)$$

$$\mathbb{P}_{\zeta_{Di}^d} \left\{ \begin{array}{l} \boldsymbol{\theta}_{un,D}^{\min} \leq \boldsymbol{A}_{P_D} \tilde{\boldsymbol{P}}_D + \boldsymbol{A}_{Q_D} \tilde{\boldsymbol{Q}}_D \leq \boldsymbol{\theta}_{un,D}^{\max} \\ \boldsymbol{V}_{un,D}^{\min} \leq \boldsymbol{B}_{P_D} \tilde{\boldsymbol{P}}_D + \boldsymbol{B}_{Q_D} \tilde{\boldsymbol{Q}}_D \leq \boldsymbol{V}_{un,D}^{\max} \end{array} \right\} \geq 1 - \delta_D^V, \quad (32)$$

$$\mathbb{P}_{\zeta_{Di}^d} \left\{ \begin{array}{l} (\tilde{P}_{Dl}^{\max})^2 + (\tilde{Q}_{Dl}^{\max})^2 \leq (S_{flow,Dl}^{\max})^2, \forall Dl \in Line_D \\ (\tilde{P}_{Dl}^{\min})^2 + (\tilde{Q}_{Dl}^{\min})^2 \leq (S_{flow,Dl}^{\max})^2, \forall Dl \in Line_D \end{array} \right\} \geq 1 - \delta_D^L. \quad (33)$$

*5) Boundary Power Constraints*

$$(P_{Tb}^B - \Delta p_b)^2 + (Q_{Tb}^B)^2 \leq (S_b)^2, \Delta p_b \in [-\Delta p_b^{-,\max}, \Delta p_b^{+,\max}]. \quad (34a)$$

$$(P_{Db}^B - \Delta p_b)^2 + (Q_{Db}^B)^2 \leq (S_b)^2, \Delta p_b \in [-\Delta p_b^{-,\max}, \Delta p_b^{+,\max}]. \quad (34b)$$

Following the above analysis, constraints (34a) and (34b) ensure that the boundary powers of TPS and ADN during SFR stay within the capacity limits of the boundary bus.

Moreover, to make the model easier to solve, the nonlinear terms in the form of $P^2 + Q^2 \leq S^2$ in constraints (7), (8), (33), and (34) can be linearized as follows [14], where $K$ is the number of linearized segments.

$$-S \leq \cos(k\frac{\pi}{K})P + \sin(k\frac{\pi}{K})Q \leq S, k = 1 \ldots K. \quad (35)$$

*D. Objective Function of the CFC-SED Model*

The objective function of the CDC-SED model shown in (36a) is to minimize the total operating cost of TPS and all ADNs. The operating costs of TPS include the generation and reserve cost of thermal units $\text{Cost}_T^G$, the expected spillage cost of DWF $\text{Cost}_T^G$, and the generation and reserve cost of ES $\text{Cost}_T^G$, which are calculated as (36b), wherein $\mathbb{E}_{\tilde{P}_{Tw}^W}$ is the expected operator of spillage cost because $\tilde{P}_{Tw}^W$ is uncertain. The operating costs in the $b^{th}$ ADN are similarly calculated as (36c), where $\text{Cost}_{D,b}^G, \text{Cost}_{D,b}^{PV}, \text{Cost}_{D,b}^E$ represent the generation and reserve cost of distributed generators, expected spillage cost of DPV, and the generation and reserve cost of ES, respectively, and $\mathbb{E}_{\tilde{P}_{Dj}^{PV}}$ also represents the similar expected operator with respect to the random variable $\tilde{P}_{Dj}^{PV}$.

$$\min \text{Cost}_T^G + \text{Cost}_T^W + \text{Cost}_T^E + \sum_{b \in N^B} (\text{Cost}_{D,b}^G + \text{Cost}_{D,b}^{PV} + \text{Cost}_{D,b}^E). \quad (36a)$$

$$\text{Cost}_T^G = \sum_{Tg \in G_T} C_{Tg}(P_{Tg}) + C_{Tg}^u R_{Tg}^u + C_{Tg}^d R_{Tg}^d$$

$$\text{Cost}_T^W = \sum_{Tw \in W_T} \mathbb{E}_{\tilde{P}_{Tw}^W}[C_{Tw}^W(\tilde{P}_{Tw}^W - P_{Tw}^W)] \quad . \quad (36b)$$

$$\text{Cost}_T^E = \sum_{Te \in E_T} \{C_{Te}^{loss} P_{Te}^{loos} + C_{Te}^{Eu} R_{Te}^{Eu} + C_{Te}^{Ed} R_{Te}^{Ed}\}$$

$$\text{Cost}_{D,b}^G = \sum_{Dg \in G_{D,b}} C_{Dg}(P_{Dg}) + C_{Dg}^u R_{Dg}^u + C_{Dg}^d R_{Dg}^d$$

$$\text{Cost}_{D,b}^{PV} = \sum_{Dj \in PV_{D,b}} \mathbb{E}_{\tilde{P}_{Dj}^{PV}}[C_{Dj}^{PV}(\tilde{P}_{Dj}^{PV} - P_{Dj}^{PV})] \quad . \quad (36c)$$

$$\text{Cost}_{D,b}^E = \sum_{De \in E_{D,b}} \{C_{De}^{loss} P_{De}^{loos} + C_{De}^{Eu} R_{De}^{Eu} + C_{De}^{Ed} R_{De}^{Ed}\}$$

### E. Model Summary and Classification

To comply with physical characteristics that TPS and ADNs are managed by different operators and facilitate the design of distributed algorithms, the dispatch models belonging to the transmission and distribution system as well as the coupling constraints for ITD's cooperative operation are summarized in the two following sections.

*1) Dispatch Model Belonging to TPS and ADN*

| Objective function | TPS | ADN |
|---|---|---|
| | $\text{Cost}_T^G + \text{Cost}_T^W + \text{Cost}_T^E$ | $\text{Cost}_{D,b}^G + \text{Cost}_{D,b}^{PV} + \text{Cost}_{D,b}^{PV}$ |
| The constraints in base point case | (1), (3)–(4), (8), | (2), (5)–(7), (8) |
| The constraints under uncertain disturbance | (19)–(21), (25)–(27), (28)–(30), (34a) | Similar constraints to (19)–(21), (25)–(27), (31)–(33), (34b) |

*2) Coupling Constraints for ITD*

The coupling constraints between TPS and ADNs involve constraints (9), (18), and (24), and the chance constraint (24) can be transformed into a deterministic constraint (37) by quantile principal [18], where $(\zeta_T)^{\uparrow, \delta_{SFR}^{ITD}/2}, (\zeta_T)^{\downarrow, \delta_{SFR}^{ITD}/2}$ denote the upper/lower $\delta_{SFR}^{ITD}/2$ quantile of $\zeta_T$, respectively.

$$\begin{aligned}(\zeta_T)^{\uparrow, \delta_{SFR}^{ITD}/2} &\leq \sum_{g \in G_T} R_{Tg}^u + \sum_{b \in N_B} (R_{D,b}^u - \zeta_{D,b}^{\max}), \ \zeta_T \geq 0 \\ -\sum_{g \in G_T} R_{Tg}^d &- \sum_{b \in N_B} (R_{D,b}^d + \zeta_{D,b}^{\min}) \leq (\zeta_T)^{\downarrow, \delta_{SFR}^{ITD}/2}, \zeta_T < 0\end{aligned} \quad (37)$$

In summary, the CFC-SED model is formulated. However, since TPS and ADNs belong to different operators, the proposed model is not suitable for a centralized solution, and the nonlinear JCCs also discourage efficient solving.

## III. SOLVING METHOD

### A. Two-Layer Distributed Optimization Framework

To overcome the aforementioned problems with the CFC-SED model, this section proposes a two-layer distributed optimization framework, which is illustrated in Fig. 4. In the outer layer, the CFC-SED model is divided into several frequency-constrained stochastic ED (FC-SED) models belonging to the transmission system operator (TSO) and distribution system operators (DSOs). Furthermore, the ADMM algorithm is adopted to realize the distributed cooperative optimization between TPS and ADNs by exchanging cooperative information, which includes **the base point boundary active/reactive power, total system inertia, droop coefficient, frequency security margin, and total regulation reserve**. It is worth noting that although the outer-layer distributed optimization is implemented based on ADMM, other distributed cooperative methods, e.g., augmented Lagrangian relaxation [26] and analytical target cascading [4], can also be adopted.

In the inner layer, each FC-SED model is a chance-constrained optimization problem including JCCs, and some popular methods, such as the SAA [17] or mix-SAA [18], can be applied for the solution. In this paper, the SAA method is adopted due to its high solution accuracy [17]. The two-layer optimization framework is introduced in detail below.

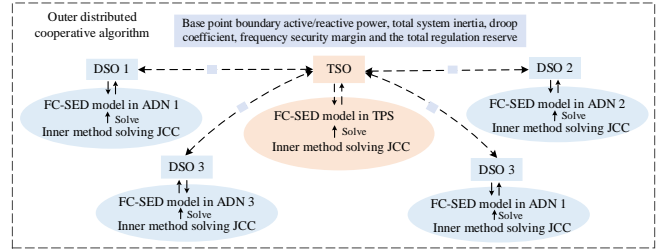

**Fig. 4.** Diagram of the distributed two-layer algorithm.

### B. ADMM Algorithm in the Outer Layer

For convenience, the mathematical abstract formulation of CFC-SED can be expressed as follows:

$$\min f_T(\boldsymbol{x}_T, \boldsymbol{y}_T) + \sum_{b \in N^B} f_{D,b}(\boldsymbol{x}_{D,b}, \boldsymbol{y}_{D,b}) , \quad (38a)$$

$$s.t. \ (\boldsymbol{x}_T, \boldsymbol{y}_T) \in \Phi_T, \ (\boldsymbol{x}_{D,b}, \boldsymbol{y}_{D,b}) \in \Phi_{D,b}, \forall b , \quad (38b)$$

$$\boldsymbol{a}_I \boldsymbol{y}_T + \boldsymbol{b}_I \boldsymbol{y}_D \geq \boldsymbol{c}_I , \quad (38c)$$

where $f_T(\boldsymbol{x}_T, \boldsymbol{y}_T), f_{D,b}(\boldsymbol{x}_{D,b}, \boldsymbol{y}_{D,b})$ represent the objective functions of TPS and the $b^{\text{th}}$ ADN, respectively; $\boldsymbol{y}_T, \boldsymbol{y}_{D,b}$ represent the corresponding cooperative variables (i.e., variables associated with coupling constraints); $\boldsymbol{y}_D$ denotes the vector involving all $\boldsymbol{y}_{D,b}$; $\boldsymbol{x}_T, \boldsymbol{x}_{D,b}$ represent the remaining intra-regional variables; and $\Phi_T, \Phi_{D,b}$ represent the operational constraints. The coupling constraints are shown in (38c), and $\boldsymbol{a}_I, \boldsymbol{b}_I, \boldsymbol{c}_I$ represent the corresponding coefficients, as discussed in Section II.

Since the coupling constraint (38c) involve inequality constraints, problem (38) cannot be solved directly by the standard ADMM algorithm. Thus, we give a decomposition method with clearer physical meaning as follows, in which problem (38) can be transformed into (39) that can be handled directly by the standard ADMM. The idea is briefly stated below.

$$\min f_T(\boldsymbol{x}_T, \boldsymbol{y}_T) + \sum_{b \in N^B} f_{D,b}(\boldsymbol{x}_{D,b}, \boldsymbol{y}_{D,b}) , \quad (39a)$$

$$s.t. \ \boldsymbol{a}_I \boldsymbol{y}_T + \boldsymbol{b}_I \boldsymbol{z}_T \geq \boldsymbol{c}_I, (\boldsymbol{x}_T, \boldsymbol{y}_T, \boldsymbol{z}_T) \in \Phi_T, (\boldsymbol{x}_{D,b}, \boldsymbol{y}_{D,b}) \in \Phi_{D,b}, (39b)$$

$$\boldsymbol{z}_T = \boldsymbol{y}_D . \quad (39c)$$

First, the coupled frequency-related constraints (i.e., constraints (18) and (37)) are incorporated into the dispatch model belonging to TPS with equivalent variable $\boldsymbol{z}_T$, which is associated with boundary power, total inertia, droop capacity, frequency safety margin, and total reserve of ADNs. Thus, problem (38) can be transformed into (39), where the coupling constraint is transformed into the equation constraint (39c) with cooperative variables $\boldsymbol{z}_T, \boldsymbol{y}_D$. Additionally, (39c) indicates that the boundary base active/reactive power and equivalent regulation capacity (i.e., total inertia, droop capacity, frequency safe margin, and SFR reserve of ADNs) expected by TSO are matched to the available capacity of ADNs, and these physical meanings are more accessible to operators.

Then, the principles of the distributed standard ADMM algorithm can be illustrated by Table I. First, the FC-SED models of TPS and ADNs are solved separately using the inner SAA method to obtain the initial value of cooperative variables in Step 1. The initialized mean value $\bar{y}^0$ of the cooperative variables between TPS and ADNs and the initialized Lagrange multipliers $\lambda_T^0, \lambda_D^0$ are calculated in Step 2.

The iterative operations are performed starting from Step 3. In Step 4, the FC-SED model with penalty term in the objective function is solved simultaneously to obtain the value of cooperative variables in the $k^{\text{th}}$ iteration, where $\lambda_{D,b}^{k-1}, \rho_{D,b}, \bar{y}_b^{k-1}$ represent the $b^{\text{th}}$ sub-vector of $\lambda_D^{k-1}, \rho_D, \bar{y}^{k-1}$. In Step 5, the mean value $\bar{y}^k$ of cooperative variables and the Lagrange multipliers $\lambda_T^k, \lambda_D^k$ in the $k^{\text{th}}$ iteration are updated. In Step 6, the convergence gap is calculated. If the iterative tolerance is satisfied, terminate the iteration and output the dispatch results of TPS and ADNs. Otherwise, return to Step 3 and iterate until the error of cooperative variables is within the tolerance.

TABLE I
PSEUDOCODE FOR ADMM ALGORITHM

**Algorithm:** Outer layer ADMM

Step 0 **Initialization:** Set iteration index $k = 0$, penalty factor $\rho_T, \rho_D$ and iterative tolerance $\varepsilon$.

Step 1 $z_T^0 = \min_{x_T, y_T, z_T} f_T(x_T, y_T), \quad a_I y_T + b_I z_T \geq c_I, (x_T, y_T, z_T) \in \Phi_T$

$y_{D,b}^0 = \min_{x_{D,b}} f_{D,b}(x_{D,b}, y_{D,b}), \quad (x_{D,b}, y_{D,b}) \in \Phi_{D,b}, \forall b$

Step 2 $\bar{y}^0 = (z_T^0 + y_D^0)/2$

$\lambda_T^0 = \rho_T(z_T^0 - \bar{y}^0), \quad \lambda_D^0 = \rho_D(y_D^0 - \bar{y}^0)$

Step 3 $k=k+1$

Step 4 $z_T^k = \min_{x_T, y_T, z_T} f_T(x_T, y_T) + \lambda_T^{k-1}(z_T - \bar{y}^{k-1}) + \rho_T \| z_T - \bar{y}^{k-1} \|_2^2$

$a_I y_T + b_I z_T \geq c_I, (x_T, y_T, z_T) \in \Phi_T$

$y_{D,b}^k = \min_{x_{D,b}, y_{D,b}} f_{D,b}(x_{D,b}, y_{D,b}) + \lambda_{D,b}^{k-1}(y_{D,b} - \bar{y}_b^{k-1}) + \rho_{D,b} \| y_{D,b} - \bar{y}_b^{k-1} \|_2^2$

$(x_{D,b}, y_{D,b}) \in \Phi_{D,b}, \forall b$

Step 5 $\bar{y}^k = (z_T^k + y_D^k)/2$

$\lambda_T^k = \rho_T(z_T^k - \bar{y}^k), \quad \lambda_D^k = \rho_D(y_D^k - \bar{y}^k)$

Step 6 $gap_T = \| z_T^k - \bar{y}^k \|_2^2, \quad gap_D = \| y_D^k - \bar{y}^k \|_2^2$

Step 7 **If** $gap_T \leq \varepsilon$ and $gap_D \leq \varepsilon$
Terminate ADMM and output the dispatch results of TPS and ADN.
**else**
Return to Step 3.
**end**

*C. Inner-Layer SAA Method and Tractability Process*

In this paper, the SAA method is adopted in the inner layer to transform each nonlinear FC-SED model into a tractable mixed-integer linear programming model. Detailed information on this process is shown in [17].

However, due to the binary indicator variables associated with sampling scenarios in the transformed mixed-integer linear programming model, the convergence of ADMM cannot be guaranteed. Inspired by [5], a tractable iterative solution method for the CFC-SED model is shown in Table II. First, by fixing binary indicator variables, ADMM is used to solve the continuous CFC-SED model to obtain the optimal value of the cooperative variable. Then, the FC-SED models of TPS and ADNs are solved independently with the above optimal cooperative variable to obtain new values of the binary indicator variables. This process is iterated, if the values of the binary indicator variables in the two iterations remain unchanged, the iteration stops, which has been validated in [5].

TABLE II
PSEUDOCODE FOR A TRACTABLE ITERATIVE ALGORITHM

**Tractable iterative algorithm**

**Initialization:** Set iteration index $t = 1$. Initialize the values of cooperative variables (by referring to steps 1 and 2 in Table I). The FC-SED models are solved independently by TSO and DSOs to obtain the optimal value of binary indicator variables $I_T^l$ and $I_{D,b}^l$ associated with sampling scenarios, respectively.

**Iteration process:**
a) Perform the ADMM algorithm in Table I with the binary indicator variable fixed to $I_T^t$ and $I_{D,b}^t$. Since the problem becomes linear and convex, the convergence of ADMM is guaranteed. Then, the optimal solution of cooperative variables can be obtained.
b) Set $t = t + 1$. Solve the FC-SED models of TPS and ADNs to obtain the optimal values of binary indicator variables $I_T^t$ and $I_{D,b}^t$, respectively.

**Iteration stopping criteria:**
If $I_T^t$ is the same as $I_T^{t-1}$, and $I_{D,b}^t$ is the same as $I_{D,b}^{t-1}$, iteration terminates, and the dispatch result is output. Otherwise, go back to iteration step a).

In summary, the distributed optimal coordination of ITD can be performed effectively via the above two-layer optimization framework, which will be further verified in the case studies.

IV. CASE STUDY

*A. Simulation Settings*

In this section, the proposed CFC-SED model and the two-layer optimization algorithm are tested in the T30-D2, T118-D9, and T300-D10 systems with a resolution of 15 min. The T30-D2 system indicates a 30-bus TPS connected with two 33-bus ADNs, where three 50 MW DWFs and three 5 MW/20 MWh ESs are connected to TPS, and four 10 MW DPVs and 2 MW/2 MWh ESs are connected to each ADN. The T118-D9 system indicates a 118-bus TPS connected with nine 69-bus ADNs, where six 300 MW DWFs and 30 MW/30 MWh ESs are connected to TPS, and four 5 MW DPVs and 2 MW/2 MWh ESs are connected to each ADN. The T300-D10 system indicates a 300-bus TPS connected with ten 69-bus ADNs. The network parameters and generation parameters of thermal units and distributed generators are derived from the IEEE test system in MTPOWER, and the charging/discharging efficiency values of the ESs are all set to 0.90/0.95, respectively. The regulation parameters of thermal units, DWFs, DPVs, and ESs are shown in [12]. In addition, the threshold values of the maximum RoCoF, the maximum frequency deviation, and the steady-state frequency are set to 0.5 Hz/s, 0.5 Hz, and 0.3 Hz, respectively. The maximum and minimum voltage thresholds are set to 1.05/0.95 p.u. The significance level $\delta^V$ is set to 0 for enhanced voltage security from a conservative perspective, and the significance levels $\delta^{IBR}$, $\delta_{SFR}^{ITD}$, $\delta^L$ are all set to 0.05.

To test the validity of the model, the active/reactive demand data and the renewable energy data from the California ISO are applied in the simulation. The forecasting errors of active demands and renewable generations are assumed to follow a beta distribution [27], and the corresponding parameters are calculated based on historical data. The sampling scenarios for SAA (500 sampling scenarios by default) associated with uncertainty are generated using Monte Carlo simulation, and the penalty factor for ADMM is set to 5.

The independent FC-SED (IFC-SED) is set as the comparison, in which a frequency-constrained stochastic ED model regarding either the TPS or an ADN is performed, ignoring the coordination between TPS and ADNs.

The codes are implemented on a computer with an Intel Core i7-11700 CPU and 16 GB RAM and solved with MATLAB and CPLEX 12.10.0, and the frequency dynamic response is verified in SIMULINK.

### B. Security Verification of the Proposed CFC-SED Model

#### 1) Mitigating System Frequency

As analyzed in Section II, the bidirectional regulation support between TPS and ADN is achieved in the CFC-SED. Obviously, the significant difference between CFC-SED and IFC-SED occurs when the regulation resources and regulation capacities are limited in TPS but are abundant in ADN (e.g., a large number of DPVs are installed), so we select and analyze this representative scenario below.

To verify the frequency security of the entire ITD system under different operating conditions when the regulation capacities in TPS are limited, four possible disturbance cases that may occur in ITD are shown in Table III, where the absolute values of disturbance in cases 1–4 are 30% of the net load of TPS and ADN. Then, the corresponding dynamic frequency performance indices, i.e., RoCoF, maximum frequency deviation (MFD), and quasi-steady frequency deviation (QFD), of the system under the dispatch results of CFC-SED and IFC-SED of TPS are shown in Table IV. The dynamic frequency response curves in case 1 are demonstrated in Fig. 5.

TABLE III
POSSIBLE DISTURBANCE CASES IN ITD

| Case | Disturbance direction | Case | Disturbance direction |
|---|---|---|---|
| Case 1 | $\zeta_T>0, \zeta_{D,b}>0$ | Case 2 | $\zeta_T<0, \zeta_{D,b}>0$ |
| Case 3 | $\zeta_T>0, \zeta_{D,b}<0$ | Case 4 | $\zeta_T<0, \zeta_{D,b}<0$ |

TABLE IV
DYNAMIC FREQUENCY INDICES OF THE ITD SYSTEM (UNIT: HZ)

| Model | IFC-SED regarding TPS | | | CFC-SED | | |
|---|---|---|---|---|---|---|
| Index | RoCoF | MFD | QFD | RoCoF | MFD | QFD |
| Case 1 | −0.487 | −0.504 | −0.310 | −0.470 | −0.468 | −0.295 |
| Case 2 | −0.395 | −0.409 | −0.251 | −0.381 | −0.380 | −0.239 |
| Case 3 | 0.395 | 0.409 | 0.251 | 0.381 | 0.380 | 0.239 |
| Case 4 | 0.487 | 0.504 | 0.310 | 0.470 | 0.468 | 0.295 |

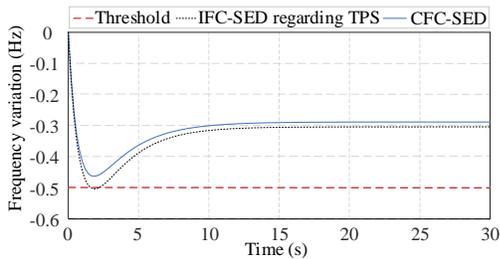

**Fig. 5.** Frequency dynamic response under disturbances of case 1 in the T30-D2 system.

As shown in Table IV and Fig. 5, if the *IFC-SED model regarding the TPS* is performed that ignores the regulation capability of ADNs, the system frequency security cannot always be guaranteed. Furthermore, frequency deviation will cross the threshold when the disturbance in case 1 or case 4 occurs, which might cause a higher operational risk and even trigger protection devices. In contrast, the CFC-SED model has more satisfactory dynamic frequency performance in all cases because the frequency regulation resources in ADNs and TPS are fully utilized and the dispatchable resources in ITDs are coordinated better, as mentioned in Section II.

In addition, the allocation of PFR-related inertia and droop capabilities and the SFR-related up/down reserve between TPS and ADNs in the dispatch result of CFC-SED are shown in Fig. 6a, indicating that CFC-SED fully exploits the frequency regulation potential of ADNs. Fig. 6b shows the base point boundary power and actual boundary power under disturbance, indicating that when the disturbance of TPS in Case 1 is positive, the boundary power from ADN to TPS increases, i.e., ADN provides upward regulation support to TPS. Furthermore, when the disturbance of TPS in Case 4 is negative, the boundary power from TPS to ADN increases, i.e., ADN provides downward regulatory support to TPS.

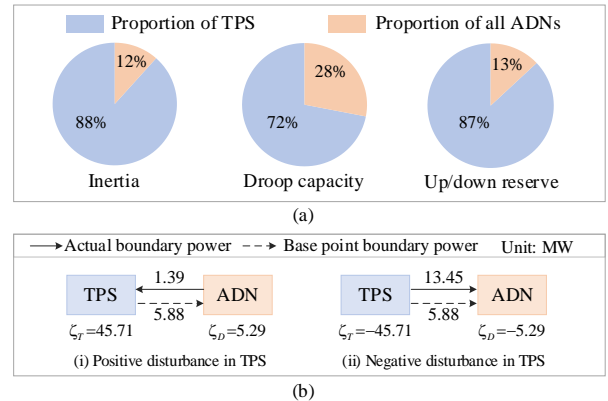

**Fig. 6.** (a) Allocation of regulation capacity between TPS and ADNs. (b) Base point and actual boundary power under disturbances.

#### 2) Mitigation of ADN's Voltage Security in the SFR Process

To verify the node voltage security of ADN in the SFR process, Monte Carlo simulation is used to generate 10 test scenarios to represent uncertain disturbance. Based on the dispatch result of the IFC-SED model regarding ADN and the CFC-SED model, the steady-state node voltage amplitudes of ADN 1 in test scenarios are shown in Fig. 7a and 7b, respectively. The *IFC-SED model regarding ADN* indicates that the base point power and reserve of all regulation units in ADN 1 are dispatched by DSO separately (i.e., ADN 1 does not participate in coordination), but the boundary power is the actual boundary power provided by TPS in actual operation.

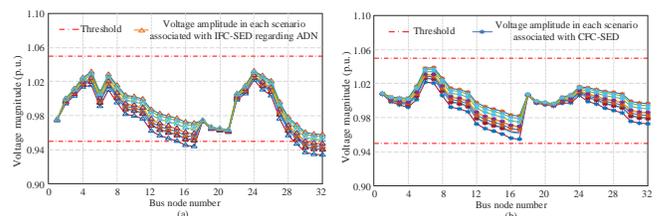

**Fig. 7.** The voltage amplitude of ADN1 in the T30-D2 system associated with (a) the IFC-SED regarding the ADN and (b) the proposed CFC-SED model.

Due to the mismatch between the expected boundary power of ADN and the actual boundary power provided by TPS, the voltage amplitudes of nodes 15, 16, and 17 are below the minimum thresholds in some scenarios in Fig. 7a. Furthermore, the voltage amplitudes of nodes 29–32 are below the minimum thresholds in most scenarios in Fig. 7a, which will cause load shedding and even trigger blackout accidents. In contrast, the CFC-SED model can ensure that the node voltage amplitudes are within the threshold in various disturbance scenarios because it can perform optimal security dispatch on the premise of cooperating with TPS and ADNs.

*C. Verification of the Two-Layer Optimization Framework*

The proposed two-layer distributed optimization framework is used to solve the CFC-SED model in the T118-D9 and T300-D10 systems, and the calculation time, iteration times, and optimal errors are shown in Table V, where the *optimal error* denotes the relative error of the objective function compared to the centralized solution. As shown in Table V, the CFC-SED model can be solved within about 0.2% optimal error. The reported calculation time can be further reduced if larger optimal errors are allowed.

TABLE V
COMPUTATIONAL PERFORMANCE OF THE PROPOSED TWO-LAYER ALGORITHM

| System | Optimal error (%) | Iteration numbers | Calculation time (s) |
|---|---|---|---|
| T118-D9 | 0.10 | 10 | 157.11 |
| T300-D10 | 0.23 | 6 | 353.37 |

V. CONCLUSIONS

In this paper, a CFC-SED model for ITD is proposed, where the base point power and regulation reserve of all dispatchable resources as well as the control parameter of IBRs in ITD are jointly optimized for optimal economy and safety. By constructing tailor-made frequency security constraints and network security constraints, TPS and ADN can deliver base point power bidirectionally and provide frequency regulation support bidirectionally, which are beneficial to enhance the regulation ability and operational safety of the ITD system. In addition, a distributed optimization framework and algorithm is proposed to solve CFC-SED. Simulation results show that the proposed CFC-SED model significantly improves the operational safety level compared with traditional separate dispatch, and the ITD system can cooperate efficiently through the two-layer optimization framework.

As this paper focuses on the operational security improvement of the system using the CFC-SED model, the market clearing or design issues to incentivize distributed photovoltaic to participate in such coordination is out of the scope and can be studied in future works.